# Singlet oxygen luminescence detection with a fiber-coupled superconducting nanowire single-photon detector

Nathan R. Gemmell,[1] Aongus McCarthy,[1] Baochang Liu,[2] Michael G. Tanner,[1] Sander N. Dorenbos,[3] Valery Zwiller,[3] Michael S. Patterson,[2] Gerald S. Buller,[1] Brian C. Wilson,[4] Robert H. Hadfield[1,5,*]

[1] *Scottish Universities Physics Alliance and School of Engineering and Physical Sciences, Heriot-Watt University, Edinburgh, EH14 4AS, United Kingdom*
[2] *Juravinski Cancer Centre and McMaster University, Hamilton, Ontario, Canada*
[3] *Kalvi Institute of Nanoscience, Delft University of Technology, 2628 CJ Delft, The Netherlands*
[4] *Department of Medical Biophysics, Ontario Cancer Institute & University of Toronto, Toronto, Ontario, Canada*
[5] *Current address: School of Engineering, University of Glasgow, Glasgow, G12 8QQ, United Kingdom*

*robert.hadfield@glasgow.ac.uk

**Abstract:** Direct monitoring of singlet oxygen ($^1O_2$) luminescence is a particularly challenging infrared photodetection problem. $^1O_2$, an excited state of the oxygen molecule, is a crucial intermediate in many biological processes. We employ a low noise superconducting nanowire single-photon detector to record $^1O_2$ luminescence at 1270 nm wavelength from a model photosensitizer (Rose Bengal) in solution. Narrow band spectral filtering and chemical quenching is used to verify the $^1O_2$ signal, and lifetime evolution with the addition of protein is studied. Furthermore, we demonstrate the detection of $^1O_2$ luminescence through a single optical fiber, a marked advance for dose monitoring in clinical treatments such as photodynamic therapy.

**1. Introduction**

Photonic technologies play an increasingly prominent role in the life sciences and medicine [1]. Many important techniques rely on time correlated single photon counting (TCSPC) [2] at visible and near infrared (NIR) wavelengths, such as fluorescence lifetime imaging [3, 4], monitoring fluorescence resonance energy transfer [5] and time-domain diffuse optical tomography [6]. These techniques commonly employ commercial off-the-shelf detectors, either large-area photomultiplier tubes (PMTs) [7] or silicon single-photon avalanche diodes (Si SPADs) [8], with spectral range up to ~1 μm wavelength [9]. The $^1\Delta_g$ excited state of molecular oxygen, commonly referred to as singlet oxygen, $^1O_2$, is important in many biological and physiological systems [10, 11]. A specific challenge for NIR single-photon detection is to measure the luminescence signature of $^1O_2$, which occurs at around 1270 nm wavelength. In the clinical domain, singlet oxygen is generated in most applications of photodynamic therapy (PDT) via photoactivation of a photosensitizing drug (Fig. 1(a)). PDT is used for the treatment of solid tumors [12, 13] and vascular disorders (e.g. age-related macular degeneration) [14]. Accurate dosimetry is needed to ensure that optimum treatment is delivered. Hence, measuring the concentration of $^1O_2$ generated during PDT, either directly or indirectly, has been the focus of considerable effort [10, 11]. Direct luminescence detection of $^1O_2$ is extremely challenging, due to its extremely high reactivity that results in very low emission probability ($\sim 10^{-8}$) and short lifetime (<<1 µs) in biological media. About a decade ago, singlet oxygen luminescence dosimetry (SOLD) became technically feasible for the first time [15], using specialized PMTs with extended NIR sensitivity. Studies of cells *in vitro* and tumors and other tissues *in vivo* [16-19] have demonstrated the validity of SOLD as a PDT dosimetry technique. Such studies have also illustrated the value of SOLD in a range of photobiological applications, from evaluating novel photosensitizers to serving as a 'gold standard' for other, indirect methods. However, the size and cost of NIR PMT detectors limits their widespread adoption, while their low quantum efficiency (<~1%) restricts their utility, in particular making it impossible to achieve single optical fiber-based light collection from biological samples and limiting their clinical utility. Although semiconductor based infrared photon counting technologies (InGaAs SPADs [9, 20]) are steadily improving, in this SOLD demonstration we turn to an emerging superconducting single-photon detector technology with exceptional performance. Superconducting nanowire single-photon detectors (SNSPDs) [21-25] offer high infrared single-photon sensitivity, combined with picosecond timing resolution, low dark count rates and free running operation. The SNSPD's low dark count rate is most useful in photon starved measurements, allowing longer acquisition times to be employed [23]. We report the first demonstration of SNSPDs for $^1O_2$ luminescence detection. We employed a fiber-coupled cavity-enhanced SNSPD with >20% practical detection efficiency at ~1300 nm [26], representing a 20–fold improvement in efficiency over NIR PMTs used to date for SOLD. The singlet oxygen signature was isolated via time-resolved and spectrally-filtered measurements performed on Rose Bengal (RB) in solution as a model photosensitizer. The free-running detector operation allows the full temporal profile of the photoluminescence signal to be analyzed. Quenching of the $^1O_2$ luminescence was demonstrated using sodium azide. Bovine serum albumen (BSA) was employed to simulate a proteinaceous biological environment; the evolution of the RB triplet and $^1O_2$ singlet-state lifetimes was successfully observed as the BSA concentration was increased. Finally, and most compellingly, an optical fiber delivery and collection scheme has been demonstrated as a breakthrough advance in SOLD for *in vivo* preclinical and clinical applications.

**2. Free space singlet oxygen luminescence monitoring**

Figure 1(b) shows a schematic of the free-space optical excitation and luminescence collection set up. A 523 nm wavelength pulsed laser source (CrystaLaser 523-500 Nd:YLF, 200 mW average power, 14 kHz repetition rate, 10 ns pulse width) passes through a 520±10 nm bandpass filter to eliminate longer-wavelength pump light, and a 0.5 OD neutral-density filter to reduce the power, before being steered through a 50 mm focal length microscope objective. A dichroic beam splitter (532 nm laser BrightLine® : reflection band 514-532 nm) diverts the beam through a second lens (25.4 mm diameter,75 mm focal length) to produce an expanded (~3 mm diameter, ~50 mW average power) collimated beam directed into the top of a 3 ml quartz cuvette filled with photosensitizer (Rose Bengal). The second plano-convex lens collimates the luminescence light and directs it through the beam splitter. A filter wheel allows selection of one of five 20 nm wide bandpass (BP) filters centered at 1210, 1240, 1270, 1310 and 1340 nm to allow sampling of the infra-red spectrum across the $^1O_2$ luminescence peak and background. Mirrors then steer the light through a 1200 nm longpass (LP) filter (Thorlabs FEL1200, cut-on wavelength 1200 nm) - to further exclude short wavelength photons - and into a 2m long single mode (9 μm core diameter) armored telecom fiber.

Coupled to the fiber is a NbTiN SNSPD with optical cavity-enhanced performance at telecom wavelengths [26], housed in a closed-cycle refrigerator (Sumitomo RDK-101D cold head with CNA-11C air-cooled compressor, operating temperature ~3 K) [22]. The SNSPD has a 70 ps full width at half maximum instrument response function with a Gaussian profile. The detection efficiency and dark count rate of the SNSPD was calibrated as a function of current bias using a gain-switched 1310 nm wavelength diode laser, depolarized and attenuated to deliver much less than one photon per pulse. The detection efficiency was assumed to vary slowly across the wavelength range of the experiment. In photon-starved experiments, the current bias on the SNSPD can be reduced to improve the signal-to-noise; with decreasing bias the dark counts drop by orders of magnitude, whereas the efficiency falls more slowly. In the free space configuration shown in Fig. 1(b), the SNSPD system detection efficiency was 15% (defined from the point where the optical signal is launched into the single mode optical fiber coupled to the SNSPD inside the cryostat) and the background (or dark) count rate was 60 counts per second (cps).

An electrical synchronization pulse from the laser gives the start signal of a time-correlated single photon counting (TCSPC) hardware module (Picoquant HydraHarp with 4 ps time binning and 20 ps FWHM timing jitter). The SNSPD output is amplified by a room temperature amplifier chain (3 dB, roll off 580 MHz) and acts as the stop channel. With the laser running at a constant pulse repetition rate of 14 kHz, the accumulation of many detector events allows generation of histograms (Fig. 1(c)) relating the time between a laser pulse and a subsequent detection event.

Since the singlet oxygen is generated by energy transfer from the photosensitizer triplet state, it has been demonstrated [15] that the concentration of $^1O_2$ at a time, $t$, following a short illumination pulse at time $t=0$ is given by:

$$[^1O_2](t) = N\sigma[S_0]\Phi_D \frac{\tau_D}{\tau_T - \tau_D}\left[\exp\left(\frac{-t}{\tau_T}\right) - \exp\left(\frac{-t}{\tau_D}\right)\right], \quad (1)$$

where $N$ is the number of photons in the pulse incident on the sample, $[S_0]$ is the ground-state photosensitizer concentration, $\sigma$ is the photosensitizer ground-state absorption cross-section, $\Phi_D$ is the $^1O_2$ quantum yield of the photosensitizer, $\tau_T$ is the photosensitizer triplet-state lifetime, and $\tau_D$ is the lifetime of $^1O_2$. Thus, after subtracting the background signal due to other sources of near-infrared emission, the histogram with the 1270 nm filter should fit Eq. (1) [15], whereas the histograms at the other wavelengths should have different form (with faint $^1O_2$ luminescence contributions in the 1240 and 1310 nm bands).

Fig. 1(c) shows representative histograms acquired from RB in each of the 5 wavelength bands. A strong initial peak (duration ~1 μs) is present in all cases, due to short lifetime broadband fluorescence from the sample and leakage of a small fraction of the laser pump light into the detector channel. However, a distinct longer-lived second peak with a broad maximum at ~3.5 μs is observed only with the 1270 nm bandpass filter in place: this is the signature of singlet oxygen luminescence. Its dependence on the RB concentration is shown in Fig. 2(a). After initial fluorescence peak and background subtraction, a luminescence photon count rate of 134 counts per second (cps) was achieved with 100 μg/ml RB. The background (dark) count rate was ~60 cps, and the detection efficiency was ~15%. Equation (1) was least-squares fitted to the time-resolved histograms, after removal of the initial fluorescence peak (using $A=N\sigma[S_0]\Phi_D$, $\tau_T$, $\tau_D$, and a constant offset, $C$, as fitting parameters), giving values of $\tau_T = 2.3\pm0.3$ μs and $\tau_D = 3.0\pm0.3$ μs. This $^1O_2$ lifetime is consistent with typical literature values in an aqueous, biomolecule-free environment [15,18].

Sodium azide, $NaN_3$, a known and highly specific singlet oxygen quencher, has been used in previous studies in solution and in cells with various photosensitizers to confirm that the 1270 nm signal comes from $^1O_2$ radiative decay [15] and not, for example, from photosensitizer triplet-state phosphorescence or other NIR background. Fig. 2(b) shows near complete loss of this signal with addition of 0.1 ml of 2 M $NaN_3$ to 3 ml of 0.257 μM RB.

As a further test of the origin of the 1270 nm signal, Fig. 3 shows the evolution of the fitted lifetimes with the addition of bovine serum albumen (BSA) to the photosensitizer, maintaining a constant RB concentration. Four samples were prepared, each containing 0.026 μM RB with 0, 2, 4, and 5.4 μM BSA, respectively. The resulting lifetimes are shown as a function of the BSA concentration in the inset of Fig. 3. As expected, the RB triplet state lifetime increases markedly with binding to the protein, likely due to shielding of the photosensitizer from the diffusion of oxygen molecules with which the triplet state can react. The decrease in $^1O_2$ lifetime has also been seen in previous studies and was attributed to quenching by the protein [15].

## 3. Optical fiber based singlet oxygen luminescence detection

Overall, the above results robustly confirm that the $^1O_2$ luminescence signal is being detected reliably with the SNSPD in the setup of Fig. 1(b), where the emitted light from the sample is first collected by free-space optics, filtered and directed into the single-mode fiber coupled to the detector. Whilst these results indicate exceptionally good signal-to-noise characteristics, this alone would not represent a paradigm shift in SOLD. The system redesign shown in Fig. 4(a), where light is delivered and collected *directly* at the sample via an optical fiber, enables a decisive step beyond previous SOLD demonstrations. Instead of the lenses and dichroic beamsplitter used in the above setup (Fig. 1(b)), two separate delivery and collection fibers are employed (Fig. 4(a)). The laser beam is launched into a collimation package attached to a 62.5 μm core diameter multimode graded index fiber. A collimation package on the other end of the fiber then produces an expanded beam (3 mm diameter) that is directed into the cuvette. A separate collimator/focuser collected the emitted light into 9 μm core diameter single-mode fiber. The light is then sent through the bandpass filters, before being coupled into the telecom fiber connected to the SNSPD. The TCSPC instrument was employed as before. A free-space section in the collection path enables the filter wheel to be employed as before for spectral selection, and the filtered luminescence signal is then directed to the fiber-coupled SNSPD. To further emulate potential *in vivo* (and indeed clinical) PDT treatments, the excitation light was also delivered via an optical fiber. The resulting time-correlated signal in each wavelength channel in Fig. 4(b) shows a weak but definitive signal in the 1270 nm window, with the same temporal features evident as in Fig. 3. Due to the poor signal-to-noise ratio, an accurate analysis of the lifetimes (such as that reported in Section 2) could not be performed . However, the loss of signal upon addition of NaN$_3$ (Fig. 4(c)) demonstrates that this is indeed the $^1O_2$ luminescence.

Experimental challenges in this configuration included optimization of the pump light delivery and alignment of the light collection scheme. While about 60% (~30 mW) of the free-space pump power was delivered to the sample, the collected signal was approximately 2 orders of magnitude lower than with free-space collection, requiring long acquisition times (~30-60 min) to obtain a reliable measurement. In order to improve the signal-to-noise we reduced the current bias on the SNSPD [23], operating at 5% detection efficiency and a background (dark) count rate of <10 cps. Under these conditions, the system detection rate of the nominal $^1O_2$ signal was estimated to be ~0.6 cps. There is considerable scope for improving this setup to drive down the acquisition time. Thus, the efficiency of the luminescence light collection could be increased considerably by using a larger-diameter fiber with high numerical aperture. In-fiber filtering could be used to eliminate the losses in the free-space gap: this would require using several fibers, for example placed in a ring around a central delivery fiber, as has been used in NIR Raman spectroscopy *in vivo* [27]. Parallel signal collection using multiple detectors in a single package would then be an attractive option. Improving the match between the excitation wavelength and the photosensitiser absorption spectrum will also proportionally increase the luminescence signal: in the present experiments, the RB absorption coefficient at the available 523 nm laser wavelength is only ~30% of the peak value at

~558 nm. Finally, the SNSPD technology evaluated here is itself advancing rapidly [24] and next-generation detectors with near 100% efficiency are under development [28].

**4. Conclusion**

In summary, we have demonstrated for the first time the feasibility of detecting singlet oxygen ($^1O_2$) luminescence using a superconducting nanowire single photon detector (SNSPD). In our initial experiments free-space light collection allows the luminescence to be coupled into the single-mode optical fiber connected to the detector, with intervening macroscopic bandpass filters to sample the NIR spectrum, enabling the broad background signal to be subtracted. The signal response to the addition of a singlet oxygen quencher (NaN$_3$) and of protein (BSA), and the ability to make high-resolution time-resolved measurements give a high level of confidence that the signal is due to $^1O_2$ luminescence emission. This is critical, since there are many confounding factors that may give misleading NIR signals, especially in cells and tissues [18]. While only a single model photosensitizer (Rose Bengal, RB) has been used in these first proof-of-principle experiments, the findings should be applicable to any $^1O_2$-generating (i.e. Type II) photodynamic drug. Excitingly, we have successfully extended SOLD for the first time to single optical fiber luminescence collection. The signal strength is then substantially reduced compared to the free-space configuration, but there are numerous technical improvements that could substantially address this loss, both in the optical design [27] and in the detector performance [28, 29]. While the possibility of fiberoptic or lightguide collection of the luminescence has been investigated by others, these efforts have either used large diameter probes [30] and/or have not been in the context of time-correlated single photon counting [31]. As shown recently, the latter is critical for accurate SOLD measurements, because of unknown and varying background contributions to the NIR signal [32]. The importance of this advance lies in the potential for greatly widening the applications of SOLD to encompass effective *in vivo* SOLD monitoring even in a clinical context: for example, fiberoptic-based detection would allow $^1O_2$ measurements in minimally-invasive endoscopic and intraoperative treatments, which are commonly used in photodynamic therapy of solid tumors. Furthermore, one can envisage placing the fiberoptic tip directly (interstitially) into the tumor or adjacent normal tissues to monitor the $^1O_2$ generation at critical sites, either to ensure that adequate PDT dose is delivered (e.g. to the base of the tumor) and/or to reduce the risk of collateral damage to neighboring critical normal tissues (e.g. in the brain or, in the case of PDT treatment of prostate cancer, the rectal wall). Finally, $^1O_2$ luminescence imaging could be possible using 2D SNSPD arrays [33, 34]. This would reveal the heterogeneity of $^1O_2$ generation within the target tissue, which is important to ensure complete response: the potential value of this concept has been demonstrated using the conventional NIR PMT detector with raster scanning of the laser beam [17], but only very low resolution images were possible in a reasonable time, since a significant per-pixel dwell time is required to collect the time-resolved signal. To conclude, we believe that the demonstration presented here, using advanced NIR detector technology to enable fiber-based SOLD for the first time, is a significant step forward for $^1O_2$ luminescence detection, and will have a profound impact in applications such as PDT.

**Acknowledgments**

The authors thank Dr Colin Rickman and Professor Rory Duncan at Heriot-Watt University for access to sample preparation facilities. RHH and BCW gratefully acknowledge the award of a Distinguished Visitor grant from the Scottish Universities Physics Alliance (SUPA), with matching support from the Heriot-Watt University Life Sciences Interface Research Theme. RHH and GSB acknowledge support from the EPSRC, UK (Grant award EP/F048041/1). RHH acknowledges a Royal Society University Research Fellowship. BCW and MSP acknowledge support from the Canadian Cancer Society Research Institute. VZ thanks FOM (the Netherlands) for support.

**Figures**

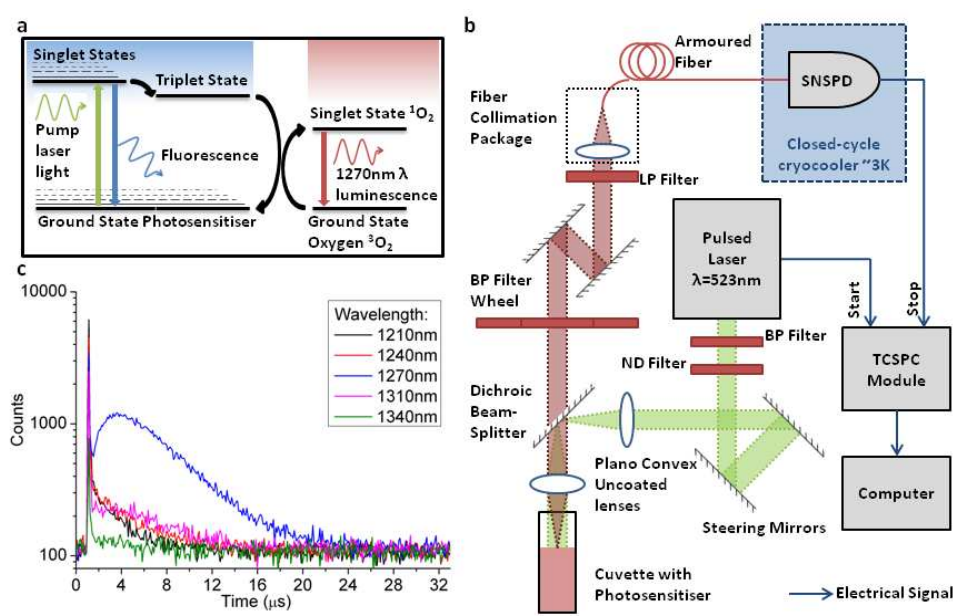

**Fig. 1 (a)** Jablonski (energy level) diagram for singlet oxygen ($^1O_2$) generation in photodynamic therapy (PDT). $^1O_2$, the primary cytotoxic species in most PDT applications, is generated by energy transfer to triplet ground-state oxygen from the excited triplet state of a photosensitizer molecule: the latter is generated from the excited singlet state that results from light absorption by the ground-state photosensitizer. $^1O_2$ can return to the ground state ($^3O_2$) with the emission of an infrared photon at around 1270 nm wavelength. However, due to the very high reaction rates of $^1O_2$ with biomolecules, this is an extremely rare event (~1 in $10^8$) in cells, and the lifetime is very short (<<1 µs). (**b**) Set-up for free space-coupled singlet oxygen luminescence detection. Acronyms used: Superconducting Nanowire Single Photon Detector (SNSPD); Long Pass (LP); Band Pass (BP); Neutral Density (ND); Time Correlated Single Photon Counting (TCSPC). (**c**) Representative TCSPC histograms from Rose Bengal solution (250 µg/ml, 0.257 µM) with bandpass filters (20 nm spectral width) centered at 1210, 1240, 1270, 1310 and 1340 nm [10 min acquisition time, 0.1024 µs bin size]. Short-lived fluorescence (occurring within the first µs) is present at all wavelengths studied; the key signature of $^1O_2$ is the onset and decay observed only at 1270 nm.

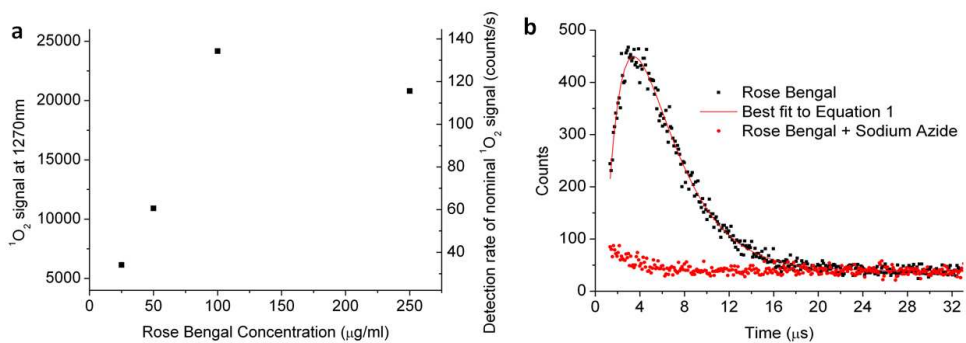

**Fig. 2 (a)** Total counts within 3 min histograms (after fluorescence peak and background subtraction), using 1270 nm band pass filtering, for increasing RB concentration. The reduction at the highest concentration is likely due to much higher attenuation of the excitation light by the increased opacity of the photosensitizer sample. **(b)** TCSPC histograms recorded from Rose Bengal (0.257 μM) with the 1270 nm bandpass filter, before and after addition of 2 M sodium azide (3 min acquisition time, 0.1024 μs bin size). The first 1.3 μs corresponding to the initial fluorescence peak is ignored. The red curve is the least-squares fit to Eq.(1), for lifetimes $\tau_T = 2.3\pm0.3$ μs and $\tau_D = 3.0\pm0.3$ μs, taking into account a constant offset due to background counts ($C$ = 39 counts per bin).

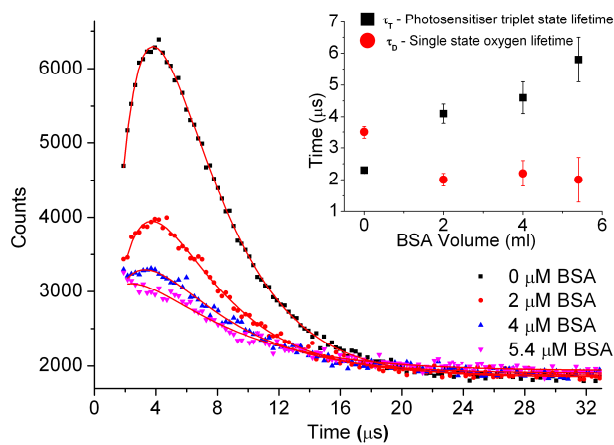

**Fig. 3** The effect of increasing BSA concentration on the Rose Bengal histograms (1270 nm filter, 60 min acquisition, 0.256 μs bin size). **Inset:** fitted lifetimes as a function of BSA concentration. The error bars represent the standard errors in the fits.

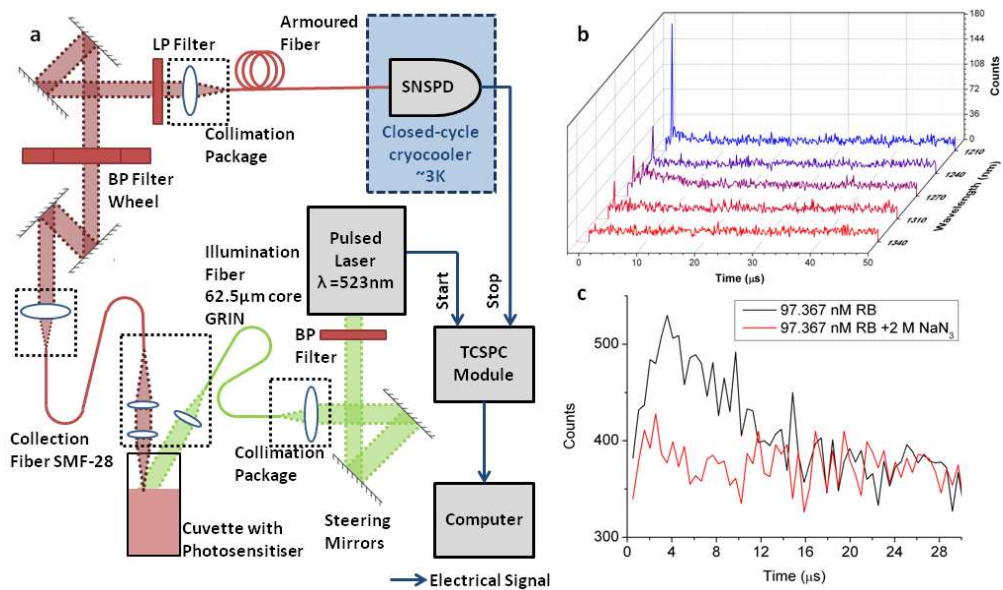

**Fig. 4** Fiber-based delivery and collection for singlet oxygen luminescence detection. (**a**) Schematic of the experimental set up. (**b**) Luminescence time histograms in RB (97.367 nM) for the different bandpass filters (30 min acquisition time, 0.128 μs bin size), (**c**) Quenching with sodium azide [60 min acquisition time, 0.512 μs bin size].